\documentclass[a4paper,UKenglish,cleveref, autoref, thm-restate]{lipics-v2021}

\usepackage{bussproofs}
\usepackage{relsize}

\usepackage{etoolbox}
\makeatletter
\patchcmd{\@verbatim}
  {\verbatim@font}
  {\verbatim@font\small}
  {}{}
\makeatother

\title{A formal proof of modal completeness for provability logic}

\author{Marco Maggesi}{University of Florence, Italy \and \url{https://www.math.unifi.it/~maggesi/} }{marco.maggesi@unifi.it}{https://orcid.org/0000-0003-4380-7691}{Supported by
the Italian Ministry of University and Research and by INdAM-GNSAGA}

\author{Cosimo Perini Brogi}{University of Genoa, Italy \and \url{https://logicosimo.gitlab.io}}{perinibrogi@dima.unige.it}{https://orcid.org/0000-0001-7883-5727}{}

\authorrunning{M. Maggesi and C. Perini Brogi} 

\Copyright{John Q. Public and Joan R. Public} 

\ccsdesc[500]{Theory of computation~Higher order logic}
\ccsdesc[500]{Theory of computation~Modal and temporal logics}
\ccsdesc[300]{Theory of computation~Automated reasoning}

\keywords{Provability Logic, Higher-Order Logic, Mechanized Mathematics, HOL Light Theorem Prover}

\relatedversion{} 

\supplement{The implementation described in this paper is hosted in the subdirectory~\texttt{GL} of the distribution of \href{https://www.cl.cam.ac.uk/~jrh13/hol-light/}{HOL Light theorem prover}.}
\supplementdetails[
  swhid={swh:1:dir:288139a46b5250ca678d928a823173f4f69c8594}]{HOL Light source code, Jenuary 2021}
{https://github.com/jrh13/hol-light/} 

\nolinenumbers 

\hideLIPIcs  

\EventEditors{John Q. Open and Joan R. Access}
\EventNoEds{2}
\EventLongTitle{42nd Conference on Very Important Topics (CVIT 2016)}
\EventShortTitle{CVIT 2016}
\EventAcronym{CVIT}
\EventYear{2016}
\EventDate{December 24--27, 2016}
\EventLocation{Little Whinging, United Kingdom}
\EventLogo{}
\SeriesVolume{42}
\ArticleNo{23}

\begin{document}
\maketitle

\begin{abstract}
This work presents a formalized proof of modal completeness for G\"odel-L\"ob provability logic (GL) in the HOL Light theorem prover.
We describe the code we developed, and discuss some details of our implementation, focusing on our choices in structuring proofs which make essential use of the tools of HOL Light and which differ in part from the standard strategies found in main textbooks covering the topic in an informal setting.
Moreover, we propose a reflection on our own experience in using this specific theorem prover for this formalization task, with an analysis of pros and cons of reasoning \emph{within} and \emph{about} the formal system for GL we implemented in our code.
\end{abstract}

\section*{Introduction}

In this paper we wish to report on our results and general experience in using HOL Light theorem prover to formally verify some properties of G\"odel-L\"ob provability logic (GL).

Our work starts with a deep embedding of the syntax of propositional modal logic together with the corresponding relational semantics.
Next, we introduce the traditional axiomatic calculus $\mathbb{GL}$ and prove the validity of the system w.r.t.~irreflexive transitive finite frames.

After doing that, the most interesting part of our work begins with the proof of a number of lemmas \emph{in} $\mathbb{GL}$ that are necessary for our main goal, namely the development of a formal proof of modal completeness for the system.

In order to achieve that, we had to formally verify a series of preliminary lemmas and constructions involving the behaviour of syntactical objects used in the standard proof of the completeness theorem. These unavoidable steps are very often only proof-sketched in wide-adopted textbooks in logic, for they mainly involve ``standard'' reasoning \emph{within} the proof system we are dealing with. But when we are working in a formal setting like we did with HOL Light, experience reveals that it is generally more convenient to adopt a different line of reasoning and to make a smart use of our formal tools, so that we can succeed in developing alternative (or simpler) proofs, still totally verified by the computer.

In other terms: in order to give a formal proof of the completeness theorem for $\mathbb{GL}$, that is
\begin{theorem}
For any formula $A$, $\mathbb{GL}\vdash A$ iff $A$ is true in any irreflexive transitive frame.
\end{theorem}
we needed to split down the main goal into several subgoals -- dealing with both the object- and  the meta-level -- which might seem trivial in informal reasoning. However, when carefully checked,  they revealed unexpected formal subtleties, which guided us towards different proof-strategies where HOL Light infrastructure played a more active role.

A more detailed account of these aspects is given in the main body of the present work, but we can briefly summarise the present paper as follows:
\begin{itemize}
\item In Section \ref{synsem}, we introduce the basic ingredients of our development, namely the formal counterparts of the syntax and relational semantics for provability logic, along with some lemmas and general definitions which are useful to handle the implementation of these objects in a uniform way, i.e.~without the restriction to the modal system we are interested in. The formalization constitutes large part of the file \verb|modal.ml|;
\item In Section \ref{axiom}, we formally define the axiomatic calculus $\mathbb{GL}$, and prove in a neat way the validity lemma for this system. Moreover, we give formal proofs of several lemmas \emph{in} $\mathbb{GL}$ ($\mathbb{GL}$-lemmas, for short), whose majority is in fact common to all normal modal logics, so that our proofs might be re-used in subsequent implementations of different systems. This corresponds to contents of our code in \verb|gl.ml|;
\item Finally, in Section \ref{compl} we introduce all the necessary to give our formal proof of modal completeness of $\mathbb{GL}$, starting with the definition of maximal consistent \emph{lists} of formulae. In order to prove their syntactic properties -- and, in particular, the extension lemma for consistent lists of formulae to maximal consistent lists -- we use the $\mathbb{GL}$-lemmas we have mentioned in the previous section, and, at the same time, we propose proof-strategies which highlight the formal tools provided by HOL Light -- or, more informally, which apply more conveniently higher-order reasoning. Therefore, the proof we are proposing in this section follows the standard lines presented in e.g.~\cite{boolos1995logic} -- from extension lemma to modal completeness via truth lemma -- but adopt tools, results, and techniques which are specific to the theorem prover we used, in line with a clear philosophical attitude in computer-aided proof development. These results, together with a simple decision procedure for $\mathbb{GL}$, are gathered in the file \verb|completeness.ml|. 
\end{itemize}

We care to stress that our formalization does not tweak any original HOL Light tools, and it is therefore ``foundationally safe''. Moreover, since we used just that original formal infrastructure, our results can be easily translated into another theorem prover belonging to the HOL family -- or, more generally, endowed with the same automation toolbox.

In any case, our code is integrated into the current HOL Light distribution, and it is freely available from there.\footnote{See "Supplementary Material" on the first page of this paper.}

Before presenting our results, in the forthcoming subsections of this introduction we provide the reader with some background material both on provability logic, and on formal theorem proving in HOL Light: further information about modalities and HOL Light can be found in \cite{goldblatt2006mathematical} and \cite{harrisontutorial}, respectively.

\subsection*{Developments of Provability Logic}
The origins of provability logic date back to a short paper  \cite{godel1933} of G\"odel's where propositions about provability are formalized by means of a unary operator $\mathtt{B}$ with the aim of giving a classical reading of intuitionistic logic.

The resulting system corresponds to the logic $\mathbb{S}4$, and the proposition $\mathtt{B} p$ has to be interpreted as `$p$ is \emph{informally} provable' -- as claimed by G\"odel himself. This implies that $\mathbb{S}4$ can be considered a provability logic lacking an appropriate semantics.

At the same time, that work opened the question of finding an adequate modal calculus for the formal properties of the provability predicate used in G\"odel's incompleteness theorems.
That problem has been settled since 1970s for many formal systems of arithmetic by means of G\"odel-L\"ob logic GL.

The corresponding axiomatic calculus $\mathbb{GL}$ consists of the axiomatic system for classical propositional logic, extended by
\begin{itemize}
\item the axiom schema \mbox{$\mathsf{K}: \Box(A\rightarrow B)\rightarrow\Box A\rightarrow\Box B$;}
\item the L\"ob schema $\mathsf{GL}$: $\Box(\Box A\rightarrow A)\rightarrow\Box A$
\item a necessitation rule \AxiomC{$A$}\UnaryInfC{$\Box A$}\DisplayProof.
\end{itemize}

The L\"ob schema is just a formal version of L\"ob's theorem which holds for a wide class of arithmetical theories satisfying the so-called Hilbert-Bernays-L\"ob (HBL) provability conditions.\footnote{These properties of the formal predicate for arithmetical provability were isolated first in \cite{hilbertbernays}.}

On the semantic side, this calculus is the logic of irreflexive transitive frames, and, moreover, it can be interpreted arithmetically in a sound and complete way. In other terms, $\mathbb{GL}$ solves the problem raised by G\"odel's paper by identifying a propositional formal system for provability in all arithmetical theories that satisfy the previously mentioned HBL conditions.

Published in 1976, Solovay's arithmetical completeness theorem \cite{solovay1976provability} is in this sense a milestone result in the fields of proof theory and modal logic: for being arithmetical complete, $\mathbb{GL}$ is capable of capturing and identifying \emph{all relevant properties of formal provability for arithmetic} in a very simple system, which is decidable and neatly characterised.

Such a deep result, however, uses in an essential way the \emph{modal} completeness of $\mathbb{GL}$: Solovay's technique basically consists of an arithmetization of a relational countermodel for a given formula that is not a theorem of $\mathbb{GL}$, from which it is possible to define an appropriate arithmetical formula that is not a theorem of the mathematical system.

In contemporary research, this is still the main strategy to prove arithmetical completeness for other modalities for provability and related concepts, in particular for interpretability logic. In spite of this, for many theories of arithmetic, including Heyting Arithmetic, this technique cannot be applied, and no alternatives are known.

Therefore, on the one hand, completeness of formal systems w.r.t.~the relevant relational semantics is still an unavoidable step in achieving the more substantial result of arithmetical completeness; on the other hand, however, the area of provability logic keeps flourishing and suggesting old and new open problems, closely related to the field of proof theory, but in fact bounded also to seeking a uniform proof-strategy to establish adequate semantics in formal theories of arithmetic having different strengths and flavours.\footnote{The reader is referred to \cite{beklemishev-visser} for a survey of open problems in provability logics. As an instance of relevant applications of this kind of formal systems to traditional investigations in proof theory see e.g.~\cite{ordinal-gl}.}  	

\subsection*{HOL Light Notation}

The HOL Light proof assistant \cite{hol-light} is based on \emph{classical} higher-order logic with polymorphic type variables and where equality is the only primitive notion. From a logical viewpoint, the formal engine defined by the \emph{term-conversions} and \emph{inference rules} underlying HOL Light is the same as that described in \cite{lambekscott}, extended by an infinity axiom and the classical characterization of Hilbert's choice operator. From a practical perspective, it is a theorem prover privileging a procedural proof style development -- i.e.~when using it, we have to solve goals by applying \emph{tactics} that reduce it to (eventually) simpler subgoals, so that the interactive aspect of proving is highlighted. Proof-terms can then be constructed by means of declarations of labels and of \emph{tacticals} that compact the proof into few lines of code evaluated by the machine. 

Logical operators -- defined in terms of equality -- and $\lambda$-abstraction are denoted by specific symbols in ASCII: for the reader's sake, we give a partial glossary to summarise them in the next table.  In the third column of the table, we also report the notations used for the object logic GL (introduced at the beginning of Section~\ref{sec:lang-sem-def}).

\medskip
\begin{tabular}{ c | c | c | l }
  Informal notation & HOL notation & GL notation & Description \\
  \hline
  $\bot$ & \verb|F| & \verb|False| & Falsity \\
  $\top$ & \verb|T| & \verb|True| & Truth \\
  $\neg p$ & \verb|~ p| & \verb|Not p| & Negation \\
  $p \wedge q$ & \verb|/\| & \verb|&&| & Conjunction \\
  $p \vee q$ & \verb|\/| & \verb$||$ & Disjunction \\
  $p \Longrightarrow q$ & \verb|==>| & \verb|-->| & Implication \\
  $p \Longleftrightarrow q$ & \verb|<=>| & \verb|<->| & Biconditional \\
  $\Box p$ &  & \verb|Box p| & Modal Operator \\
  $\forall x.\, P(x)$ & \verb|!x. P(x)| &  & Universal quantification \\
  $\exists x.\, P(x)$ & \verb|?x. P(x)| &  & Existential quantification \\
  $\lambda x.\,M(x)$ & \verb|\x. M(x)| &  & Lambda abstraction \\
\end{tabular}
\medskip

\noindent
In the following sections, we will directly state our results as theorems and definitions in the HOL Light syntax.
Note that theorems are prefixed by the turnstile symbol, as in~\verb$|- 2 + 2 = 4$.
We often report a theorem with its associated name, that is, the name of its associated OCaml constant, e.g.
\begin{verbatim}
  ADD_SYM
    |- !m n. m + n = n + m
\end{verbatim}
As expository style, we omit formal proofs at all, but meaning of definitions, lemmas, and theorems in natural language is clear after the table we have just given.

We warn the reader that the HOL Light printing mechanism omit type information completely.  However in this paper we manually add type annotations when it might be useful, or even indispensable, to avoid ambiguity -- including the case of our main results \verb|COMPLETENESS_THEOREM| and \verb|COMPLETENESS_THEOREM_GEN|.

As already told in the introduction, our contribution is now part of the HOL Light distribution.
The reader interested in performing these results in her machine -- and perhaps build further formalization on top of it -- can run our code with the command
\begin{verbatim}
  loadt "GL/make.ml";;
\end{verbatim}
at the HOL Light prompt.

\section{Basics of Modal Logic}\label{synsem}

As we stated previously, we deal with a logic that extends classical propositional reasoning by means of a single modal operator which is intended to capture the abstract properties of the provability predicate for arithmetic.

To reason about and within this logic, we have to ``teach'' HOL Light -- our meta-language -- how to identify it, starting with its syntax -- the object-language -- and semantics -- the interpretation of this very object-language.

We want to keep everything neat and clean from a foundational perspective, therefore we will define \emph{both} the object-language and its interpretation with no relation to the HOL Light environment. In other terms: our formulae and operators are \emph{real} syntactic objects which we keep distinct from their semantic counterpart -- and from the logical operators of the theorem prover too.

\subsection{Language and Semantics Defined}
\label{sec:lang-sem-def}

Let us start by fixing the propositional modal language we will use throughout the present work. We consider \emph{all} classical propositional operators -- conjunction, disjunction, implication, equivalence, negation, along with the 0-ary symbols $\top$ and $\bot$ -- and we add a modal unary connective $\Box$. The starting point is, as usual, a denumerable infinite set of propositional atoms $a_{0},a_{1},\cdots$. 
Accordingly, formulae of this language will have one of the following form $$ a\;\, |\;\, A\wedge B\,\, |\;\, A \vee B\;\, |\;\, A\rightarrow B\;\, |\;\, A\leftrightarrow B\;\,|\;\, \neg A\;\, |\;\, \top\;\, |\;\, \bot\;\,|\;\, \Box A\;\,.$$
The following code extends the HOL system with an the \textbf{inductive type of formulae} up to the \textbf{atoms} -- which we identify with the denumerable type of strings -- by using the above \textbf{connectives}:
\begin{verbatim}
  let form_INDUCT,form_RECURSION = define_type
    "form = False
          | True
          | Atom string
          | Not form
          | && form form
          | || form form
          | --> form form
          | <-> form form
          | Box form";;
\end{verbatim}
Next, we turn to the semantics for our modal language. We use \textbf{relational models} -- aka Kripke models\footnote{See \cite{kripkestoria} for the historical development of this notion.} -- which consist of directed graphs labelled with formulae.
More formally, a \textbf{Kripke frame} is made of a nonempty set `of possible worlds' \verb|W|, together with a binary relation \verb$R$ on \verb$W$. To this, we add a valuation function \verb$V$ which assigns to each atom of our language and each world \verb$w$ in \verb$W$ a boolean value.
This is extended to a \textbf{forcing relation} \texttt{holds}, defined recursively on the structure of the input formula \verb$p$, that computes the truth-value of \verb$p$ in a specific world \verb$w$:
\begin{verbatim}
  let holds = new_recursive_definition form_RECURSION
    `(holds WR V False (w:W) <=> F) /\
   	 (holds WR V True w <=> T) /\
   	 (holds WR V (Atom s) w <=> V s w) /\
   	 (holds WR V (Not p) w <=> ~(holds WR V p w)) /\
   	 (holds WR V (p && q) w <=> holds WR V p w /\ holds WR V q w) /\
   	 (holds WR V (p || q) w <=> holds WR V p w \/ holds WR V q w) /\
   	 (holds WR V (p --> q) w <=> holds WR V p w ==> holds WR V q w) /\
   	 (holds WR V (p <-> q) w <=> holds WR V p w <=> holds WR V q w) /\
   	 (holds WR V (Box p) w <=> !u. u IN FST WR /\ SND WR w u ==> holds WR V p u)`;;
\end{verbatim}
Then, the \textbf{validity} of a formula \verb|p| with respect to a frame \verb|(W,R)|, and a class of frames \verb|L|, denoted respectively \verb|holds_in (W,R) p| and \verb$L |= p$, are
\begin{verbatim}
  let holds_in = new_definition
    `holds_in (W,R) p <=> !V w. w IN W ==> holds (W,R) V p w`;;
\end{verbatim}
\begin{verbatim}
  let valid = new_definition
    `L |= p <=> !f. L f ==> holds_in f p`;;
\end{verbatim}
The above definitions are already presented in Chapter 20 of Harrison's HOL Light Tutorial \cite{harrisontutorial}.

\subsection{Frames for GL}
\label{sec:frames-GL}
For carrying out our formalization, we are interested in the logic of the (nonempty) frames whose underlying relation $R$ is \textbf{transitive} and conversely well-founded -- aka \textbf{Noetherian} -- on the corresponding set of possible worlds; in other terms, we want to study the modal tautologies in models based on an accessibility relation $R$ on $W$ such that
\begin{itemize}
\item if $xRy$ and $yRz$, then $xRz$; and
\item for no $X\subseteq W$ which is nonempty, there are infinite $R$-chains $x_{0}Rx_{1}Rx_{2}\cdots$.
\end{itemize}
In HOL Light, \verb|WF R| states that \verb|R| is a well-founded relation, so that we express the latter condition as \verb|WF(\x y. R y x)|.
Here we see a recurrent motif in logic: defining a system from the semantic perspective requires non-trivial tools from the foundational point of view, for, to express the second condition, a first-order language is not enough. However, that is not an issue here, since our underlying system is natively higher order:

\begin{verbatim}
  let TRANSNT = new_definition
    `TRANSNT (W,R) <=>
     ~(W = {}) /\
     (!x y:W. R x y ==> x IN W /\ y IN W) /\
     (!x y z:W. x IN W /\ y IN W /\ z IN W /\ R x y /\ R y z ==> R x z) /\
     WF(\x y. R y x)`;;
\end{verbatim}
From a theoretical point of view, moreover, the question has no deep consequences as we can characterize this class of frames by using a \emph{propositional} language extended by a modal operator $\Box$ that satisfies the following axiom -- usually called \emph{G\"odel-L\"ob axiom schema}:
$$\mathsf{GL}:\quad\Box(\Box A\rightarrow A)\rightarrow\Box A.$$
Here is the formal version of our claim:
\begin{verbatim}
  LOB_IMP_TRANSNT
    |- !W R. (!x y:W. R x y ==> x IN W /\ y IN W) /\
             (!p. holds_in (W,R) (Box(Box p --> p) --> Box p))
             ==> (!x y z. x IN W /\ y IN W /\ z IN W /\ R x y /\ R y z ==> R x z) /\
                 WF (\x y. R y x)
\end{verbatim}

\begin{verbatim}
  TRANSNT_IMP_LOB
    |- !W R. (!x y:W. R x y ==> x IN W /\ y IN W) /\
             (!x y z. x IN W /\ y IN W /\ z IN W /\ R x y /\ R y z ==> R x z) /\
             WF (\x y. R y x)
             ==> (!p. holds_in (W,R) (Box(Box p --> p) --> Box p))
\end{verbatim}

\begin{verbatim}
  TRANSNT_EQ_LOB
    |- !W R. (!x y:W. R x y ==> x IN W /\ y IN W)
             ==> ((!x y z. x IN W /\ y IN W /\ z IN W /\ R x y /\ R y z
                           ==> R x z) /\
                  WF (\x y. R y x) <=>
                  (!p. holds_in (W,R) (Box(Box p --> p) --> Box p)))
\end{verbatim}
Note that the main claim is proven in both its directions by using both a defined tactic introduced in \cite{harrisontutorial}, and the principle of well-founded induction as implemented in HOL Light.

By using this preliminary result, we could say that the frame property of being transitive and Noetherian can be captured by G\"odel-L\"ob modal axiom, without recurring to a higher-order language. Nevertheless, that class of frames is not particularly informative from a logical point of view: in particular, we have no information about the cardinality of $W$, so that it might be possible to find very exotic models for which the axiom holds, with no real interest yet.

We would like, in fact, to sharpen this result by identifying a smaller class of frames which is eventually more informative and easier to reason about. To this aim we note that by definition of Notherianness, our $R$ cannot be reflexive -- otherwise $xRxRx\cdots$ would give us an infinite $R$-chain. This is not enough: the frames we want to investigate are precisely those whose $W$ is \textbf{finite}, and whose $R$ is both \textbf{irreflexive} and \textbf{transitive}:

\begin{verbatim}
  let ITF = new_definition
    `ITF (W:W->bool,R:W->W->bool) <=>
     ~(W = {}) /\
     (!x y:W. R x y ==> x IN W /\ y IN W) /\
     FINITE W /\
     (!x. x IN W ==> ~R x x) /\
     (!x y z. x IN W /\ y IN W /\ z IN W /\ R x y /\ R y z ==> R x z)`;;
\end{verbatim}
Now it is easy to see that \texttt{ITF} is a subclass of \texttt{TRANSNT}:
\begin{verbatim}
  ITF_NT
    |- !W R:W->W->bool. ITF(W,R) ==> TRANSNT(W,R)
\end{verbatim}
That will be the class of frames whose logic we are now going to define syntactically.

\section{Axiomatizing GL}\label{axiom}

As stated before, we want to identify the logical system generating all the modal tautologies for transitive Noetherian frames; more precisely, we want to isolate the \emph{generators} of the modal tautologies in the subclass of transitive Noetherian frames which are finite, transitive, and irreflexive.

When dealing with the very notion of tautology -- or \emph{theoremhood}, discarding the complexity or structural aspects of \emph{derivability} in a formal system -- it is convenient to focus on axiomatic calculi. The calculus we are dealing with here is usually denoted by $\mathbb{GL}$.

It is clear from the definition of the forcing relation that for classical operators any axiomatization of propositional classical logic will do the job.
Here, we adopt a basic system in which only $\rightarrow$ and $\bot$ are primitive -- from the axiomatic perspective -- and all the remaining classical connectives are defined by axiom schemas and by the inference rule of Modus Ponens imposing their standard behaviour.\footnote{This is essentially the calculus introduced by Wajsberg in 1939, and, independently by Church in 1956.}

To this classical engine we add
\begin{itemize}
\item the axiom schema $\mathsf{K}$:\quad $\Box(A\rightarrow B)\rightarrow\Box A\rightarrow\Box B$;
\item the axiom schema $\mathsf{GL}$:\quad $\Box(\Box A\rightarrow A)\rightarrow\Box A$;
\item the necessitation rule $\mathsf{NR}$: \ \AxiomC{$A$}\RightLabel{$\mathsmaller{\mathsf{NR}}$}\UnaryInfC{$\Box A$}\DisplayProof ,
\end{itemize}
where $A,B$ are generic formulae (not simply atoms).
Then, here is the complete definition of the \textbf{axiom system} $\mathbb{GL}$.
The set of axioms is encoded via the inductive predicate \verb|GLaxiom|:
\begin{verbatim}
  let GLaxiom_RULES,GLaxiom_INDUCT,GLaxiom_CASES = new_inductive_definition
    `(!p q. GLaxiom (p --> (q --> p))) /\
     (!p q r. GLaxiom ((p --> q --> r) --> (p --> q) --> (p --> r))) /\
     (!p. GLaxiom (((p --> False) --> False) --> p)) /\
     (!p q. GLaxiom ((p <-> q) --> p --> q)) /\
     (!p q. GLaxiom ((p <-> q) --> q --> p)) /\
     (!p q. GLaxiom ((p --> q) --> (q --> p) --> (p <-> q))) /\
     GLaxiom (True <-> False --> False) /\
     (!p. GLaxiom (Not p <-> p --> False)) /\
     (!p q. GLaxiom (p && q <-> (p --> q --> False) --> False)) /\
     (!p q. GLaxiom (p || q <-> Not(Not p && Not q))) /\
     (!p q. GLaxiom (Box (p --> q) --> Box p --> Box q)) /\
     (!p. GLaxiom (Box (Box p --> p) --> Box p))`;;
\end{verbatim}
The judgment $\mathbb{GL} \vdash A$, denoted \verb$|-- A$ in the machine code, is also inductively defined in the expected way:
\begin{verbatim}
  let GLproves_RULES,GLproves_INDUCT,GLproves_CASES = new_inductive_definition
    `(!p. GLaxiom p ==> |-- p) /\
     (!p q. |-- (p --> q) /\ |-- p ==> |-- q) /\
     (!p. |-- p ==> |-- (Box p))`;;
\end{verbatim}

\subsection{$\mathbb{GL}$-lemmas}

As usual, $\mathbb{GL}\vdash A$ denotes the existence of a derivation of $A$ from the axioms of $\mathbb{GL}$; we could also define a notion of derivability from a set of assumptions just by tweaking the previous definitions in order to handle the specific limitations on $\mathsf{NR}$ -- so that the deduction theorem would hold \cite{fittinginhandbook} -- but this would be inessential to our intents.

What matters for the subsequent development is to derive further 
$\mathbb{GL}$-lemmas with the help of the automation in HOL Light. In accordance with this aim, we denoted the classical axioms and rules of the system as the propositional schemas used by Harrison in the file \verb|Arithmetic/derived.ml| of the HOL Light standard distribution~\cite{hol-light} -- where, in fact, many of our lemmas relying on the propositional calculus only are already proven there w.r.t.~an axiomatic system for first-order classical logic; our further lemmas involving modal reasoning are denoted with names that are commonly used in informal presentations.

The code in \texttt{gl.ml} mainly consists then of the formalized proofs of those lemmas \emph{in} $\mathbb{GL}$ that are useful for the formalized results we present in the next section.
This file might be thought as a ``kernel'' for further experiments in reasoning about axiomatic calculi by using HOL Light. The lemmas we proved are, indeed, standard tautologies of classical propositional logic, along with specific theorems of minimal modal logic and its extension for transitive frames -- i.e.~of the systems $\mathbb{K}$ and $\mathbb{K}4$ \cite{popkorn1994first} --, so that by applying minor changes in basic definitions, they are -- so to speak -- take-away proof-terms for extensions of that very minimal system within the realm of normal modal logics.

More precisely, we have given, whenever it was useful, a ``natural deduction'' characterization of classical operators both in terms of an implicit (or internal) deduction -- and in that case we named the lemma with the suffix \verb|_th| --, such as
\begin{verbatim}
  GL_modusponens_th
    |- !p q. |-- ((p --> q) && p --> q)
\end{verbatim}
and as a derived rule of the axiomatic system mimicking the behaviour of the connective in Gentzen's formalism, e.g.,
\begin{verbatim}
  GL_and_elim
    |- !p q r. |-- (r --> p && q) ==> |-- (r --> q)  /\ |-- (r --> p)
\end{verbatim}
We had to prove about 120 such results of varying degree of difficulty.  We believe that this file is well worth the effort of its development, for two main reasons to be considered -- along with the just mentioned fact that they provide a (not so) minimal set of internal lemmas which can be moved to different axiomatic calculi at, basically, no cost. 

Indeed, on the one hand, these lemmas simplify the subsequent formal proofs involving consistent lists of formulae since they let us work formally within the scope of $\vdash$, so that we can rearrange subgoals according to their most useful equivalent form by applying the appropriate $\mathbb{GL}$-lemma(s).

On the other hand, the endeavour of giving formal proofs of these lemmas of the calculus $\mathbb{GL}$ has been important for checking how much our proof-assistant is ``friendly'' and efficient in performing this specific task. 

As it is known, any axiomatic system fits very well an investigation involving the notion of \emph{theoremhood} for a specific logic, but its lack of naturalness w.r.t.~the practice of developing informal proofs makes it an unsatisfactory model for the notion of \emph{deducibility}. In more practical terms: developing a formal proof of a theorem in an axiomatic system \emph{by pencil and paper} can be a dull and uninformative task. 

After approaching this very same objective by using HOL Light, we feel like expressing mixed opinions on the general experience. In most cases, relying on the automation tools of this specific proof assistant did indeed save our time and resources when trying to give a formal proof in $\mathbb{GL}$. Nevertheless, there has been a number of $\mathbb{GL}$-lemmas for proving which those automation tools did not revealed useful at all. In those cases, actually, we had to perform a tentative search of the specific instances of axioms from which deriving the lemmas,\footnote{The HOL Light tactics for first-order reasoning \texttt{MESON} and \texttt{METIS} were unable, for instance, of instantiating autonomously the obvious middle formula for the transitivity of an implication, or even the specific formulae of a schema to apply to the goal in order to rewrite it.} so that interactive proving them had advantages as well as traditional instruments of everyday mathematicians.

Just to stress the general point: it is clearly possible -- and actually useful in general -- to rely on the resources of HOL Light to develop formal proofs both \emph{about} and \emph{within} an axiomatic calculus for a specific logic, in particular when the lemmas of the object system have relevance or practical utility for mechanizing (meta-)results on it; however, these very resources -- and, as far as we can see, the tools of any other general proof assistant -- do not look peculiarly satisfactory for pursuing investigations on derivability within axiomatic systems.

\subsection{Soundness Lemma}

At this point, we can prove that $\mathbb{GL}$ is \textbf{sound} -- i.e.~every formula derivable in the calculus is a tautology in the class of irreflexive transitive frames.
This is obtained by simply unfolding the relevant definitions and applying theorems \verb|TRANSNT_EQ_LOB| and \verb|ITF_NT| of Section~\ref{sec:frames-GL}:


\begin{verbatim}
  GL_TRANSNT_VALID
    |- !p. (|-- p) ==> TRANSNT:(W->bool)#(W->W->bool)->bool |= p
\end{verbatim}

\begin{verbatim}
  GL_ITF_VALID
    |- !p. |-- p ==> ITF:(W->bool)#(W->W->bool)->bool |= p
\end{verbatim}
From this, we get a model-theoretic proof of \textbf{consistency for the calculus}
\begin{verbatim}
  GL_consistent
    ~ |-- False
\end{verbatim}
This having exhausted the contents of \verb|gl.ml|, we would move to consider now the most interesting part of our effort, namely the mechanized proof of completeness for the calculus w.r.t.~this very same class of frames. That constitutes the remaining contents of our implementation, beside the auxiliary code in \verb|misc.ml| furnishing some general results about lists of items with same type we needed to handle the subsequent constructions, but have more general utility.

\section{Completeness and Decidability}\label{compl}
When dealing with normal modal logics, it is common to develop a proof of completeness w.r.t.~relational semantics by using the so-called `canonical model method'. This can be summarized as a standard construction of countermodels made of maximal consistent sets of formulae and an appropriate accessibility relation~\cite{popkorn1994first}.

For $\mathbb{GL}$, we cannot pursue this strategy, since the logic is not compact: maximal consistent sets are (in general) infinite objects, though the notion of derivability involves only a finite set of formulae. For being not compact, we cannot therefore reduce the semantic notion of (in)coherent set of formulae to the syntactic one of (in)consistent set of formulae: when extending a consistent set of formulae to a maximal consistent one, we might end up with a \emph{syntactically} consistent set that nevertheless cannot be \emph{semantically} satisfied. 

In spite of this, it is possible to achieve a completeness result by
\begin{enumerate}
\item identifying the relevant properties of maximal consistent sets of formulae; and
\item tweaking the definitions so that those properties hold for specific consistent sets of formulae related to the formula we want to find a countermodel to.
\end{enumerate}

As a matter of fact, we can claim something more: it is possible to carry out in a \emph{very natural way} a related tweaked Lindenbaum construction to extend consistent \emph{lists} to maximal consistent ones, preserving the standard line of reasoning in completeness proofs, and, at the same time, avoiding symbolic subtleties that have no real relevance for the argument, but have the unpleasant consequence of making the formalized proof unnecessarily long, so that the implementation would sound rather pedantic -- or even dull.    

\subsection{Maximal Consistent Lists}

As just claimed, in order to make efficient use of our formal device, we will work with \emph{lists} of formulae, instead of simple sets.\footnote{See Section \ref{sec:max} below for the reasons of our choice.}
Accordingly, we define first, by recursion on the list, the operation of finite conjunction of formulae in a list:
\begin{verbatim}
  let CONJLIST = new_recursive_definition list_RECURSION
    `CONJLIST [] = True /\
     (!p X. CONJLIST (CONS p X) = if X = [] then p else p && CONJLIST X)`;;
\end{verbatim}
Afterwards, we prove some properties of this operation on lists that characterize it in terms of $\mathbb{GL}$-derivability, and which depends, basically, on most of the $\mathbb{GL}$-lemmas defined above. In particular, since $\mathbb{GL}$ is a normal modal logic -- i.e.~its modal operator distributes over implication and preserves theoremhood -- we have that our modal operator distributes over the conjunction of $X$ so that we have \verb|CONJLIST_MAP_BOX|: $$\mathbb{GL}\vdash\Box\bigwedge X\leftrightarrow\bigwedge\Box X,$$ where $\Box X$ is an abuse of notation for the list obtained by ``boxing'' each formula in $X$.

We are now able to define the notion of \textbf{consistent list of formulae} and prove the main properties of this kind of objects:

\begin{verbatim}
  let CONSISTENT = new_definition
    `CONSISTENT (l:form list) <=> ~ (|-- (Not (CONJLIST l)))`;;
\end{verbatim}
In particular, we prove that:
\begin{itemize}
\item a consistent list cannot contain either both $A$ and $\neg A$ for any formula $A$, nor $\bot$ (see theorems \verb|CONSISTENT_LEMMA|, \verb|CONSISTENT_NC|, and \verb|FALSE_IMP_NOT_CONSISTENT|, respectively);
\item for any consistent list $A$ and formula $A$, $X+A$ is consistent, or $X+\neg A$ is consistent (\verb|CONSISTENT_EM|), where $+$ denotes the usual operation of appending an element to a list.
\end{itemize}
Our \textbf{maximal consistent lists} w.r.t.~a given formula $A$ will be consistent lists that do not contain repetitions and that contain, for any subformula of $A$, that very subformula or its negation:\footnote{Here we define the set of subformulae of $A$ as the reflexive transitive closure of the set of formulae on which the main connective of $A$ operates: this way, the definition is simplified and it is easier to establish standard properties of the set of subformulae by means of general higher-order lemmas in HOL Light for the closure of a given relation.}

\begin{verbatim}
  let MAXIMAL_CONSISTENT = new_definition
    `MAXIMAL_CONSISTENT p X <=>
     CONSISTENT X /\ NOREPETITION X /\
     (!q. q SUBFORMULA p ==> MEM q X \/ MEM (Not q) X)`;;
\end{verbatim}
where \verb|X| is a list of formulae and \verb|MEM q X| is the membership relation for lists.
We then establish the main closure property of maximal consistent lists:
\begin{verbatim}
  MAXIMAL_CONSISTENT_LEMMA
    |- !p X A b. MAXIMAL_CONSISTENT p X /\
                 (!q. MEM q A ==> MEM q X) /\
                 b SUBFORMULA p /\
                 |-- (CONJLIST A --> b)
                 ==> MEM b X
\end{verbatim}
After proving some further lemmas with practical utility -- in particular, the fact that any maximal consistent list behaves like a restricted bivalent evaluation for classical connectives (\verb|MAXIMAL_CONSISTENT_MEM_NOT| and \verb|MAXIMAL_CONSISTENT_MEM_CASES|) -- we can finally define the ideal (type of counter)model we are interested in.
The type \verb|STANDARD_MODEL| consists, for a given formula $A$, of:
\begin{enumerate}
\item the set of maximal consistent lists w.r.t.~$A$ made of subformulae of $A$ or their negations, as possible worlds;
\item\label{2.} an irreflexive transitive accessibility relation $R$ such that for any subformula $\Box B$ of $A$ and any world $w$, $\Box B$ is in $w$ iff for any $x$ $R$-accessible from $w$, $B$ is in $x$;
\item an atomic valuation that gives value $1$ to $p$ in $w$ iff $p$ is a subformula of $A$.
\end{enumerate}

Here is the corresponding code:

\begin{verbatim}
  let GL_STANDARD_FRAME = new_definition
    `GL_STANDARD_FRAME p (W,R) <=>
     W = {w | MAXIMAL_CONSISTENT p w /\
              (!q. MEM q w ==> q SUBFORMULA p \/
                               (?q'. q = Not q' /\ q' SUBFORMULA p))} /\
     ITF (W,R) /\
     (!q w. Box q SUBFORMULA p /\ w IN W
            ==> (MEM (Box q) w <=> !x. R w x ==> MEM q x))`;;
\end{verbatim}

\begin{verbatim}
  let GL_STANDARD_MODEL = new_definition
    `GL_STANDARD_MODEL p (W,R) V <=>
     GL_STANDARD_FRAME p (W,R) /\
     (!a w. w IN W ==> (V a w <=> MEM (Atom a) w /\ Atom a SUBFORMULA p))`;;
\end{verbatim}

\subsection{Maximal Extensions}\label{sec:max}

What we have to do now is to show that the type \verb|GL_STANDARD_MODEL| is nonempty. We achieve this by constructing suitable maximal consistent lists of formulae from specific consistent ones.

Our original strategy differs from the presentation given in e.g.~\cite{boolos1995logic} for being closer to the standard Lindenbaum construction commonly used in proving completeness results. By doing so, we have been able to circumvent both the pure technicalities in formalizing the combinatorial argument sketched in \cite[p.79]{boolos1995logic} \emph{and} the problem -- apparently inherent to the Lindenbaum extension -- due to the non-compactness of the system, as we mentioned before.

Given a formula $A$, the key idea is to extend a consistent list of subformulae of $A$ or their negations in a step-by-step construction, by adding at each stage of the process a subformula of $A$ -- if the resulting list is consistent -- or -- otherwise -- the negation of that very subformula, after fixing an enumeration of the subformulae of $A$:

\begin{verbatim}
  EXTEND_MAXIMAL_CONSISTENT
    |- !p X.
         CONSISTENT X /\
         (!q. MEM q X ==> q SUBFORMULA p \/ (?q'. q = Not q' /\ q' SUBFORMULA p))
         ==> ?M. MAXIMAL_CONSISTENT p M /\
                 (!q. MEM q M
                      ==> q SUBFORMULA p \/
                          (?q'. q = Not q' /\ q' SUBFORMULA p)) /\
                 X SUBLIST M
\end{verbatim}
This way, we are in the pleasant condition of carrying out the construction by using the HOL Light device efficiently, and, at the same time, we do not have to worry about the non-compactness of $\mathbb{GL}$ since we are working with finite objects -- the type \texttt{list} -- by default. 

Henceforth, we see that -- under the assumption that $A$ is not a $\mathbb{GL}$-lemma -- the set of possible worlds in \verb|STANDARD_FRAME| w.r.t.~$A$ is nonempty, as required by the definition of relational structures:

\begin{verbatim}
  NONEMPTY_MAXIMAL_CONSISTENT
    |- !p. ~ |-- p
           ==> ?M. MAXIMAL_CONSISTENT p M /\
                   MEM (Not p) M /\
                   (!q. MEM q M ==> q SUBFORMULA p \/
                                    (?q'. q = Not q' /\ q' SUBFORMULA p))
\end{verbatim}
Next, we have to define an $R$ satisfying the condition \ref{2.} for a
\verb|STANDARD_FRAME|; the following does the job:

\begin{verbatim}
  let GL_STANDARD_REL = new_definition
    `GL_STANDARD_REL p w x <=>
     MAXIMAL_CONSISTENT p w /\
     (!q. MEM q w ==> q SUBFORMULA p \/ ?q'. q = Not q' /\ q' SUBFORMULA p) /\
     MAXIMAL_CONSISTENT p x /\
     (!q. MEM q x ==> q SUBFORMULA p \/ ?q'. q = Not q' /\ q' SUBFORMULA p) /\
     (!B. MEM (Box B) w ==> MEM (Box B) x /\ MEM B x) /\
     (?E. MEM (Box E) x /\ MEM (Not (Box E)) w)`;;
\end{verbatim}
Such an accessibility relation, together with the set of the specific maximal consistent lists we are dealing with, defines a structure in \verb|ITF| with the required properties:

\begin{verbatim}
  ITF_MAXIMAL_CONSISTENT
    |- !p. ~ |-- p
           ==> ITF ({M | MAXIMAL_CONSISTENT p M /\
                         (!q. MEM q M ==> q SUBFORMULA p \/
                                          ?q'. q = Not q' /\ q' SUBFORMULA p)},
                    GL_STANDARD_REL p),
\end{verbatim}

\begin{verbatim}
  ACCESSIBILITY_LEMMA
    |- !p M w q.
         ~ |-- p /\
         MAXIMAL_CONSISTENT p M /\
         (!q. MEM q M ==> q SUBFORMULA p \/ (?q'. q = Not q' /\ q' SUBFORMULA p)) /\
         MAXIMAL_CONSISTENT p w /\
         (!q. MEM q w ==> q SUBFORMULA p \/ (?q'. q = Not q' /\ q' SUBFORMULA p)) /\
         MEM (Not p) M /\
         Box q SUBFORMULA p /\
         (!x. GL_STANDARD_REL p w x ==> MEM q x)
         ==> MEM (Box q) w,
\end{verbatim}

\subsection{Truth Lemma and Completeness}

For our ideal model, it remains to reduce the semantic relation of forcing to the more tractable one of membership to the specific world. More formally, we prove -- by induction on the complexity of the subformula $B$ of $A$ -- that if $\mathbb{GL}\not\vdash A$, then for any world $w$ of the standard model, $B$ holds in $w$ iff $B$ is member of $w$:

\begin{verbatim}
  GL_truth_lemma
    |- !W R p V q.
         ~ |-- p /\
         GL_STANDARD_MODEL p (W,R) V /\
         q SUBFORMULA p
         ==> !w. w IN W ==> (MEM q w <=> holds (W,R) V q w),
\end{verbatim}
Finally, we are able to prove the main result: if $\mathbb{GL}\not\vdash A$, then the list $ [\neg A]$ is consistent, and by applying \verb|EXTEND_MAXIMAL_CONSISTENT|, we obtain a maximal consistent list $X$ w.r.t.~$A$ that extends it, so that, by applying \verb|GL_truth_lemma|, we have that $X\not\vDash A$ in our standard model. The corresponding formal proof reduces to the application of those previous results and the appropriate instantiations:

\begin{verbatim}
  COMPLETENESS_THEOREM
    |- !p. ITF:(form list->bool)#(form list->form list->bool)->bool |= p
           ==> |-- p,
\end{verbatim}
Notice that the family of frames \verb|ITF| is polymorphic, but, at this stage, our result holds only for frames on the domain \verb|form list|, as indicated by the type annotation. This is not an intrinsic limitation: the next section is devoted indeed to generalize this theorem to frames on an arbitrary domain.

As an immediate corollary, we have that the system $\mathbb{GL}$ is decidable and, in principle, we could implement a decision procedure for it in OCaml.
However, this is a difficult task -- especially if one seeks efficiency and completeness -- and it is out of the scope of the present work.
Nevertheless, we feel like offering a very rough approximation.  

We define the tactic \verb|GL_TAC| and its associated rule \verb|GL_RULE| that perform the following steps: (1) apply the completeness theorem, (2) unfold some definitions, and (3) try to solve the resulting \emph{semantic} problem using first-order reasoning.
\begin{verbatim}
  let GL_TAC : tactic =
    MATCH_MP_TAC COMPLETENESS_THEOREM THEN
    REWRITE_TAC[valid; FORALL_PAIR_THM; holds_in; holds;
                ITF; GSYM MEMBER_NOT_EMPTY] THEN
    MESON_TAC[];;
\end{verbatim}
\begin{verbatim}
  let GL_RULE tm = prove(tm, REPEAT GEN_TAC THEN GL_TAC);;
\end{verbatim}
The above naive strategy is able to prove automatically some lemmas which are common to normal modal logic, but require some effort when derived in an axiomatic system. As an example consider the following $\mathbb{GL}$-lemma:
\begin{verbatim}
  GL_box_iff_th
    |- !p q. |-- (Box (p <-> q) --> (Box p <-> Box q))  
\end{verbatim}

When developing a proof of it within the axiomatic calculus, we need to ``help'' HOL Light with by instantiating several further $\mathbb{GL}$-lemmas, so that the resulting proof-term consists of ten lines of code.
On the contrary, our rule is able to check it in few steps:
\begin{verbatim}
  # GL_RULE `!p q. |-- (Box (p <-> q) --> (Box p <-> Box q))`;;
    0..0..1..6..11..19..32..solved at 39
    0..0..1..6..11..19..32..solved at 39
  val it : thm = |- !p q. |-- (Box (p <-> q) --> (Box p <-> Box q))
\end{verbatim}
In spite of this, the automation offered by \verb|MESON| tactic is not enough when the generated semantic problem involves in an essential way the fact that our frames are finite (or Noetherian). So, for instance, our procedure is not able to prove the Löb axiom
$$\mathbb{GL} \vdash \Box(\Box A \longrightarrow A) \longrightarrow \Box A\,.$$
This suggests the need for improving \verb|GL_TAC| to handle more general contexts: in the long run, it is a likely-looking outcome of what we reached so far.



\subsection{Generalizing via Bisimulation}
\label{sec:bisim}

As we stated before, our theorem~\verb|COMPLETENESS_THEOREM| provides the modal completeness for $\mathbb{GL}$ with respect to a semantics defined using models built on the type \verb|:form list|.  It is obvious that the same result must hold whenever we consider models built on any infinite type.  To obtain a formal proof of this fact, we need to establish a \emph{correspondence} between models built on different types.  It is well-known that a good way to make rigorous such correspondence is through the notion of \emph{bisimulation} \cite{blackburn}.

In our context, given two models \verb|(W1,R1)| and \verb|(W2,R2)| sitting respectively on types \verb|:A| and \verb|:B|, each with a valuation function \verb|V1| and \verb|V2|, a \textbf{bisimulation} is a binary relation \verb|Z:A->B->bool| that relates two worlds \verb|w1:A| and \verb|w2:B| when they can \emph{simulate} each other.  The formal definition is as follows:
\begin{verbatim}
  BISIMIMULATION
    |- BISIMIMULATION (W1,R1,V1) (W2,R2,V2) Z <=>
       (!w1:A w2:B.
          Z w1 w2
          ==> w1 IN W1 /\ w2 IN W2 /\
              (!a:string. V1 a w1 <=> V2 a w2) /\
              (!w1'. R1 w1 w1'  ==> ?w2'. w2' IN W2 /\ Z w1' w2' /\ R2 w2 w2') /\
              (!w2'. R2 w2 w2' ==> ?w1'. w1' IN W1 /\ Z w1' w2' /\ R1 w1 w1'))
\end{verbatim}
Then, we say that two worlds are \emph{bisimilar} if there exists a bisimulation between them:
\begin{verbatim}
  let BISIMILAR = new_definition
    `BISIMILAR (W1,R1,V1) (W2,R2,V2) (w1:A) (w2:B) <=>
     ?Z. BISIMIMULATION (W1,R1,V1) (W2,R2,V2) Z /\ Z w1 w2`;;
\end{verbatim}
The key fact is that the semantic predicate \verb|holds| respects bisimilarity:


\begin{verbatim}
  BISIMILAR_HOLDS
    |- !W1 R1 V1 W2 R2 V2 w1:A w2:B.
         BISIMILAR (W1,R1,V1) (W2,R2,V2) w1 w2
         ==> (!p. holds (W1,R1) V1 p w1 <=> holds (W2,R2) V2 p w2)
\end{verbatim}
From this, we can proof that validity is preserved by bisimilarity.  The precise statements are the following:
\begin{verbatim}
  BISIMILAR_HOLDS_IN
    |- !W1 R1 W2 R2.
         (!V1 w1:A. ?V2 w2:B. BISIMILAR (W1,R1,V1) (W2,R2,V2) w1 w2)
         ==> (!p. holds_in (W2,R2) p ==> holds_in (W1,R1) p)
\end{verbatim}
\begin{verbatim}
  BISIMILAR_VALID
    |- !L1 L2.
         (!W1 R1 V1 w1:A.
            L1 (W1,R1) /\ w1 IN W1
            ==> ?W2 R2 V2 w2:B. L2 (W2,R2) /\
                                BISIMILAR (W1,R1,V1) (W2,R2,V2) w1 w2)
         ==> (!p. L2 |= p ==> L1 |= p)
\end{verbatim}
Finally, we can explicitly define a bisimulation between \verb|ITF|-models on the type \verb|:form list| and on any infinite type \verb|:A|.
From this, it follows at once the desired generalization of completeness for $\mathbb{GL}$:
\begin{verbatim}
  COMPLETENESS_THEOREM_GEN
    |- !p. INFINITE (:A) /\ ITF:(A->bool)#(A->A->bool)->bool |= p ==> |-- p
\end{verbatim}

\section*{Related Works}

Our formalization gives a mechanical proof of completeness for $\mathbb{GL}$ in HOL Light which sticks to the original Henkin's method for classical logic. In its standard version, its nature is synthetic and intrinsically semantic \cite{fittingmendelsohn}, and, as we stated before, it is the core of the canonical model construction for most of normal modal logic.

That very approach does not work for $\mathbb{GL}$. Nevertheless, the modified extension lemma we proved in our mechanization introduces an analytic flavour to the strategy -- for building maximal consistent lists in terms of components of the given formula non-proven in the calculus -- and shows that Henkin's idea can be applied to $\mathbb{GL}$ too modulo appropriate changes.

As far as we know, \emph{no other mechanized proof of modal completeness for} $\mathbb{GL}$ has been given before, despite there exist formalizations of similar results for several other logics, manly propositional and first-order classical and intuitionistic logic.

Formalizations of completeness for \emph{classical logic} define an established trend in interactive theorem proving since \cite{shankar}, where a Hintikka-style strategy is used to define a theoremhood checker for formulas built up by negation and disjunction only.

In fact, a very general treatment of systems for classical propositional logic is given in \cite{nipkow}. There, axiomatic calculus is investigated along with natural deduction, sequent calculus, and resolution system in Isabelle/HOL, and completeness is proven by Hintikka-style method for sequent calculus first, to be lifted then to the other formalisms by means of translations of each system into the others. Their formalization is more ambitious than ours, but, at the same time, it is focused on a very different aim.

Finally, a recent overview of meta-theoretical results for several calculi and theorem provers formalized in HOL/Isabelle is given in \cite{blanchette}, where, in any case, a different proof assistant is used towards a more general investigation unrelated to modal logics.

An objective more similar to ours is obtained in \cite{bentzen}, where the Henkin-style completeness for the system $\mathbb{S}5$ is formalized within the Lean theorem prover -- therefore, it uses a different meta-theory and works on the classical method of canonical model for normal modal logic.

\bibliography{ref}

\end{document}